\documentclass[a4paper,11pt]{article}

\usepackage{scrlfile}
\PreventPackageFromLoading{natbib}
\usepackage{Layout/pos}
\usepackage[backend=biber,style=numeric-comp,sorting=none,giveninits=true,
  maxbibnames=7,minbibnames=1,doi=true,url=false,isbn=false,eprint=true]{biblatex}
\DeclareNameAlias{default}{given-family}
\DeclareFieldFormat[article,misc]{title}{\mkbibemph{#1}}
\DeclareFieldFormat{journaltitle}{#1}
\DeclareFieldFormat{issuetitle}{#1}
\DeclareFieldFormat[article,misc]{volume}{\mkbibbold{#1}}
\DeclareFieldFormat{pages}{#1}
\DeclareFieldFormat{eid}{#1}
\DeclareFieldFormat{usera}{\mkbibparens{#1}}
\DeclareDelimFormat{finalnamedelim}{\addspace\bibstring{and}\space}
\DeclareFieldFormat{eprint:arxiv}{\href{https://arxiv.org/abs/#1}{arXiv:#1}}
\DeclareFieldFormat{doi}{\href{https://doi.org/#1}{doi:\nolinkurl{#1}}}
\AtEveryBibitem{\clearfield{eprintclass}}
\ExplSyntaxOn
\renewbibmacro*{author}{
  \printnames{author}
  \setunit{\addspace}\printfield{usera}}
\DeclareBibliographyDriver{article}{
  \usebibmacro{bibindex}
  \usebibmacro{begentry}
  \usebibmacro{author}
  \setunit{\addcomma\space}\usebibmacro{title}
  \setunit{\addcomma\space}\printfield{journaltitle}
  \setunit{\addspace}
  \iffieldundef{issuetitle}{\printfield{volume}}{\printfield{issuetitle}}
  \ifentrytype{misc}{}{\setunit{\addspace}\printtext[parens]{\printdate}}
  \setunit{\addspace}\printfield{eid}\printfield{pages}
  \setunit{\addcomma\space}
  \iffieldundef{eprint}{\printfield{doi}}{\usebibmacro{eprint}}
  \usebibmacro{finentry}}
\ExplSyntaxOff
\DeclareBibliographyAlias{misc}{article}
\usepackage{bbold}
\usepackage{lineno}
\usepackage{dcolumn}
\usepackage{slashed}
\usepackage{wrapfig}
\usepackage{adjustbox}
\usepackage{placeins}
\usepackage[font=small,labelfont=bf,justification=centering]{subcaption}
\usepackage{enumitem}
\usepackage{booktabs}
\usepackage{tikz}
\usetikzlibrary{arrows.meta}
\usepackage[nameinlink]{cleveref}
\graphicspath{{Figures/}{Layout/}}
\setlist[itemize]{leftmargin=1.35em,itemsep=2pt,topsep=4pt,parsep=0pt}
\setlist[enumerate]{leftmargin=1.6em,itemsep=2pt,topsep=4pt,parsep=0pt}
\newcommand{\Nf}{N_f}
\newcommand{\Nt}{N_\tau}
\newcommand{\bpc}{\beta_{\mathrm{pc}}}

\newcommand{\clqcd}{\texttt{CL\kern-.25em\textsuperscript{2}QCD}}
\renewcommand{\acknowledgments}{\section*{Acknowledgments}
  \vspace{-1em}
}

\title{Identifying the conformal window by chiral phase boundaries}
\ShortTitle{Identifying the conformal window by chiral phase boundaries}
\author*[a]{Jan Philipp Klinger}
\author[a,b]{Reinhold Kaiser}
\author[a,b]{Owe Philipsen}
\author[a]{Jonas Schaible}
\affiliation[a]{Institute for Theoretical Physics - Goethe University, \\
Max-von-Laue-Str. 1, 60438 Frankfurt am Main, Germany}
\affiliation[b]{John von Neumann Institute for Computing (NIC) at GSI,\\
Planckstr. 1, 64291 Darmstadt, Germany}
\emailAdd{klinger@itp.uni-frankfurt.de}
\emailAdd{kaiser@itp.uni-frankfurt.de}
\emailAdd{philipsen@itp.uni-frankfurt.de}
\emailAdd{jschaible@itp.uni-frankfurt.de}

\abstract{
Once the number of massless fermion flavours reaches a critical value $N_f^*$, many-flavour QCD enters the conformal window, where chiral symmetry remains unbroken in the vacuum. At non-zero quark mass, chiral symmetry is explicitly broken, and thermal chiral crossovers can also occur for $N_f\geq N_f^*$. Their presence alone therefore does not distinguish a QCD-like theory from a mass-deformed conformal theory.
They differ, however, in whether a thermal transition persists in the continuum chiral limit. To address this, we map out the chiral phase boundaries of unimproved staggered fermions in the bare lattice parameter space $(N_\tau,\beta,am,N_f)$, comparing $N_f=7$ and $N_f=8$. Our results indicate that the thermal chiral transition persists in the continuum chiral limit for $N_f=7$, whereas the massless continuum theory with $N_f=8$ remains chirally symmetric in the vacuum. Furthermore, we demonstrate that the pseudo-critical surface for eight flavours observed for finite quark masses is consistent with conformal hyperscaling, providing further evidence that eight-flavour QCD lies within the conformal window and suggesting $7<N_f^*\leq8$.}

\FullConference{43rd International Symposium on Lattice Field Theory (Lattice 2026)\\
26 July -- 1 August 2026\\
University of Maryland, College Park, USA}

\begin{document}
\maketitle

\section{Introduction}
\label{sec:introduction}

The perturbative two-loop beta function of an $SU(3)$ gauge theory with $\Nf$ massless fermions predicts an interacting infrared fixed point for sufficiently many flavours~\cite{BanksZaks}. Such a fixed point would govern the long-distance dynamics and render the theory infrared conformal.
Beyond perturbation theory, lattice studies provided evidence for infrared conformality, for example at $N_f=10$~\cite{Hasenfratz2023} and $N_f=12$~\cite{Lombardo_12,PetersonHasenfratz2024}. For an overview, see Ref.~\cite{KotovLombardoTrunin2021}. However, the location of the lower edge of the conformal window, $\Nf^*$, remains uncertain, with a history of widely varying estimates from just above six flavours~\cite{Iwasaki2004} to around twelve~\cite{MiuraLombardo2013} on the lattice and near thirteen in early functional-RG studies~\cite{Braun:2006jd,Braun:2009ns}. More recently, a functional-RG analysis placed the loss of spontaneous chiral symmetry breaking at $\Nf\simeq7.3$~\cite{PastorGutierrez2026}, and a lattice study of the renormalised gradient-flow coupling suggested that the eight-flavour theory may lie at the lower edge of the conformal window~\cite{Hasenfratz2022}. A complementary lattice study at $\Nf=8$ found the Roberge--Weiss transition temperature to be consistent with zero in the chiral limit~\cite{DElia2026}.

\begin{figure}[htbp]
  \centering
  \includegraphics[width=.60\linewidth]{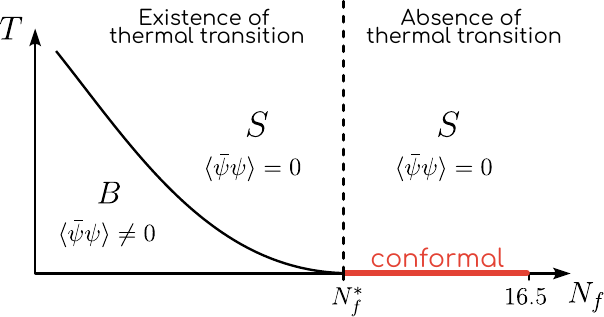}
  \caption{Schematic phase diagram of massless many-flavour QCD with chirally broken ($B$) and symmetric ($S$) phases.
The unknown $\Nf^*$ separates the QCD-like regime from the conformal window.}
  \label{fig:temperature-flavour}
\end{figure}

For $SU(3)$ gauge theory with $\Nf\geq3$ massless fundamental flavours, 't~Hooft anomaly matching and persistent-mass constraints require spontaneous chiral symmetry breaking in the confining regime~\cite{Ciambriello2024}. Figure~\ref{fig:temperature-flavour} illustrates the scenario in which chiral symmetry restoration at $T=0$ coincides with the onset of infrared conformality at $\Nf^*$. For $\Nf<\Nf^*$ there is a chiral phase transition whose temperature decreases with $\Nf$, while for $\Nf\geq\Nf^*$ the theory is chirally symmetric for all temperatures including zero. This motivates us to map out the chiral phase boundaries in the extended bare parameter space
spanned by the lattice gauge coupling $\beta=6/g^2$, the number of flavours $\Nf$,
the bare quark mass $am$, and the temporal lattice extent $\Nt$.
Extending our previous study~\cite{Klinger2026}, we compare staggered lattice QCD with $\Nf=7$ and $8$. At finite quark mass and sufficiently large $N_\tau$, both theories exhibit thermal crossovers. Our results, however, reveal a clear distinction in how these transitions approach the chiral limit.
For $\Nf=7$, the thermal transition is connected to the continuum chiral limit.
For $\Nf=8$, by contrast, the massless theory remains chirally symmetric in the continuum at all temperatures. Moreover, the corresponding pseudo-critical temperature scales with the bare quark mass in a manner consistent with conformal hyperscaling as the chiral limit is approached.

\vspace{5mm}

\begin{figure}[!htbp]
  \centering
  \begin{subfigure}[t]{.48\linewidth}
    \centering\includegraphics[width=\linewidth,height=3.5cm,keepaspectratio]{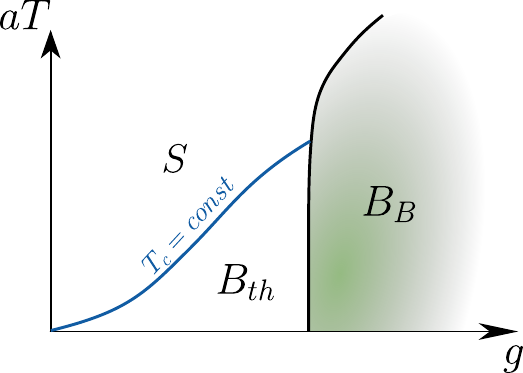}
    \caption{QCD-like theory: $N_f < N_f^*$}\label{fig:at-g-qcd}
  \end{subfigure}\hfill
  \begin{subfigure}[t]{.48\linewidth}
    \centering\includegraphics[width=\linewidth,height=3.5cm,keepaspectratio]{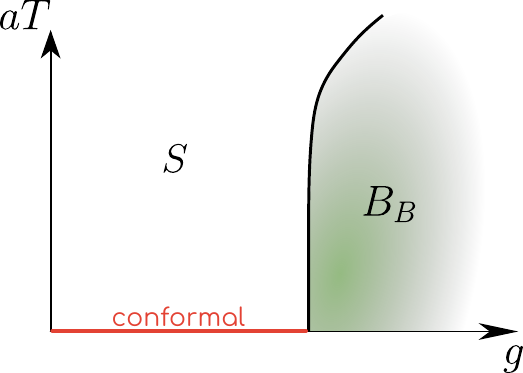}
    \caption{Conformal scenario: $N_f \geq N_f^*$}\label{fig:at-g-conformal}
  \end{subfigure}
\caption{Schematic phase structure of massless lattice QCD in the $(g,aT)$ plane.
Shading marks the chirally broken bulk phase $B_B$.
For a QCD-like theory (\subref{fig:at-g-qcd}), a thermal boundary separates the chirally broken region $B_{\rm th}$
from the symmetric region $S$ and connects to the Gaussian continuum limit $(g,aT)=(0,0)$.
For a conformal theory (\subref{fig:at-g-conformal}), the physical regime remains chirally symmetric at all temperatures.}
\label{fig:at-g}
\end{figure}

\begin{figure}[!htbp]
  \centering
  \begin{subfigure}[t]{.48\linewidth}
    \centering\includegraphics[width=\linewidth,height=3.5cm,keepaspectratio]{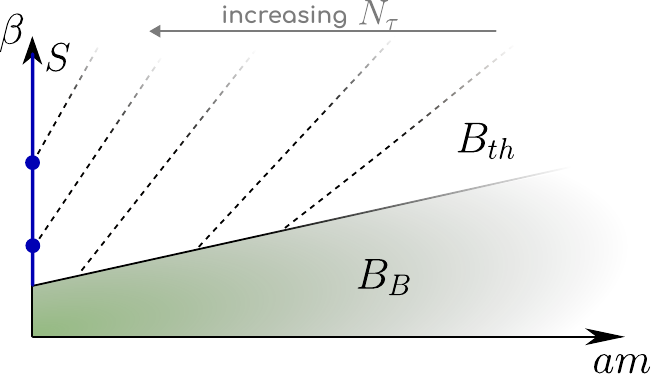}
    \caption{QCD-like theory: $N_f < N_f^*$}\label{fig:beta-am-qcd}
  \end{subfigure}\hfill
  \begin{subfigure}[t]{.48\linewidth}
    \centering\includegraphics[width=\linewidth,height=3.5cm,keepaspectratio]{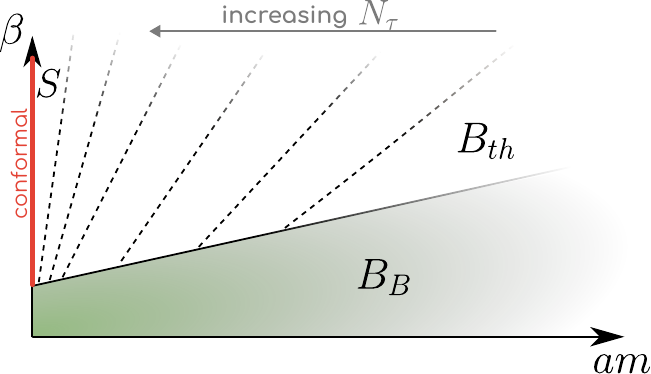}
    \caption{Conformal scenario: $N_f \geq N_f^*$}\label{fig:beta-am-conformal}
  \end{subfigure}
\caption{Schematic phase structure in the $(\beta,am)$ plane for increasing $\Nt=1/aT$, with $\beta=6/g^2$. For a QCD-like theory (\subref{fig:beta-am-qcd}), the thermal transition reaches $am=0$ above the bulk boundary for sufficiently large $\Nt$.
For a conformal theory (\subref{fig:beta-am-conformal}), the physical regime at $am=0$ remains chirally symmetric for all $\Nt$.
The chiral limits of these diagrams reproduce the massless phase structure in Figure~\ref{fig:at-g}.}
\label{fig:beta-am}
\end{figure}

\section{Lattice phase structure}
\label{sec:geometry}

To guide the interpretation of our results, we briefly summarise the lattice phase structure expected for the unimproved staggered fermion action used here in the bare parameter space $(\beta,am,\Nt,\Nf)$. Following Ref.~\cite{Klinger2026}, we discuss how the continuum phase diagram in Figure~\ref{fig:temperature-flavour} emerges upon taking the continuum and chiral limits.

Figure~\ref{fig:at-g} illustrates the expected lattice phase structure in the $(g,aT)$ plane for a massless QCD-like theory and a conformal theory at fixed $\Nf$. At sufficiently strong bare coupling, many-flavour lattice QCD develops a bulk phase $B_B$ in which lattice artefacts induce chiral symmetry breaking~\cite{Brown1992,Iwasaki2004,Klinger2026}. A bulk transition separates this region from the physical weak-coupling regime connected to the continuum. This transition persists at zero temperature, and at sufficiently large \(\Nt\), its location no longer changes. For a QCD-like theory, illustrated in Figure~\ref{fig:at-g-qcd}, a thermal boundary separates the physical weak-coupling regime into a chirally broken low-temperature region $B_{\rm th}$ and a symmetric region $S$. A non-zero continuum transition temperature $T_\chi$ requires this boundary to approach the Gaussian continuum limit, $g\to0$ and $aT_\chi=1/\Nt\to0$, while keeping $T_\chi=1/(a(g)\Nt)$ fixed.
In the conformal scenario, by contrast, the massless theory remains chirally symmetric at all temperatures, including $T=0$. Consequently, no thermally broken region $B_{\rm th}$ is expected on the physical weak-coupling side. Only the bulk boundary separates $B_B$ from the continuum-connected symmetric region $S$ for all temporal extents $N_\tau$, as shown in Figure~\ref{fig:at-g-conformal}.

 Numerical simulations necessarily run at non-zero quark mass, which explicitly breaks chiral symmetry. In this case thermal crossovers also occur for $\Nf\geq\Nf^*$. The key distinction between a QCD-like theory and a mass-deformed conformal theory is whether a thermal transition persists in the continuum chiral limit. This becomes transparent in the $(\beta,am)$ projections of the phase boundaries at different $\Nt$, shown in Figure~\ref{fig:beta-am}. For sufficiently large $\Nt$ the bulk boundary remains $\Nt$-independent at non-zero mass, whereas the thermal boundary moves towards larger $\beta$ as $\Nt$ increases.
For a QCD-like theory, the bulk transition obscures the thermal transition in the chiral limit at small $\Nt$. At sufficiently large $\Nt$, a thermal transition emerges at $am=0$ on the physical side of the bulk boundary. The pseudo-critical coupling then increases towards the continuum limit, $\beta\to\infty$ as $\Nt\to\infty$, reproducing the massless phase structure in Figure~\ref{fig:at-g-qcd}. In the conformal scenario, no thermal boundary reaches $am=0$ on the physical weak-coupling side of the bulk transition, and the massless theory remains chirally symmetric at all temperatures.

\section{Lattice setup}
\label{sec:setup}

Our simulations employ the standard Wilson gauge action with unimproved, mass-degenerate staggered fermions on $\Nt\times N_\sigma^3$ lattices. We use \clqcd~\cite{CL2QCD}, an OpenCL lattice QCD code suite, with job management and monitoring
provided by the Bash tool BaHaMAS~\cite{BaHaMAS}.
We locate the chiral transitions using the chiral condensate
$\langle\mathcal{O}\rangle=\langle\bar\psi\psi\rangle$ as a quasi-order parameter.
Its distribution is characterised by the standardised central moments
\begin{equation}
 B_n=\frac{\left\langle\left(\mathcal{O}-\langle\mathcal{O}\rangle\right)^n\right\rangle}
 {\left\langle\left(\mathcal{O}-\langle\mathcal{O}\rangle\right)^2\right\rangle^{n/2}}.
 \label{eq:central-moments}
\end{equation}
At fixed $\Nf$, $\Nt$ and $am$, we combine ensembles at neighbouring couplings using
multihistogram reweighting \cite{FerrenbergSwendsen1989} and determine the pseudo-critical coupling
from vanishing skewness, $B_3(\bpc,am)=0$. For both $\Nf=7$ and $8$, we used lattices with aspect ratio $N_\sigma/\Nt=2$ at $\Nt=10,12,16$ and $N_\sigma/\Nt\simeq1.5$ at $\Nt=14,18,22$. We assessed finite-volume effects on $\bpc$ using larger aspect ratios at selected parameter points.
\begin{figure}[htbp]
  \centering
  \begin{subfigure}[t]{.5\linewidth}
    \centering
    \makebox[\linewidth][c]{\includegraphics[width=1.10\linewidth]{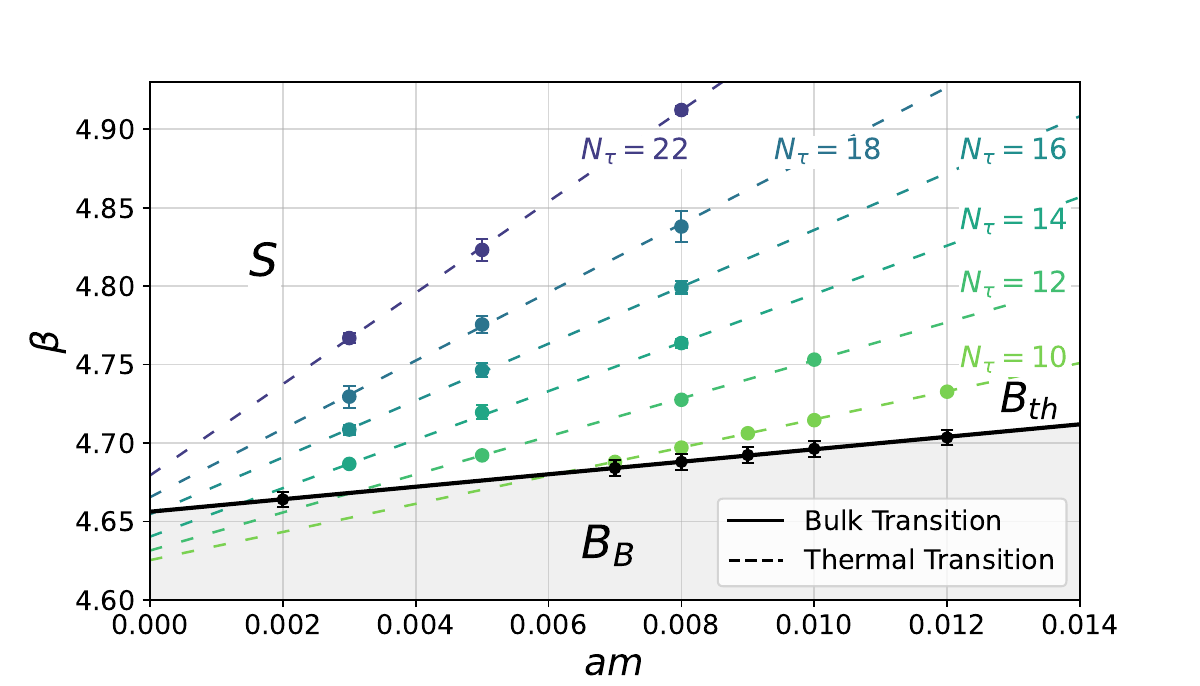}
    }
    \caption{$\Nf=7$}\label{fig:lines-seven}
  \end{subfigure}\begin{subfigure}[t]{.5\linewidth}
    \centering
    \makebox[\linewidth][c]{\includegraphics[width=1.10\linewidth]{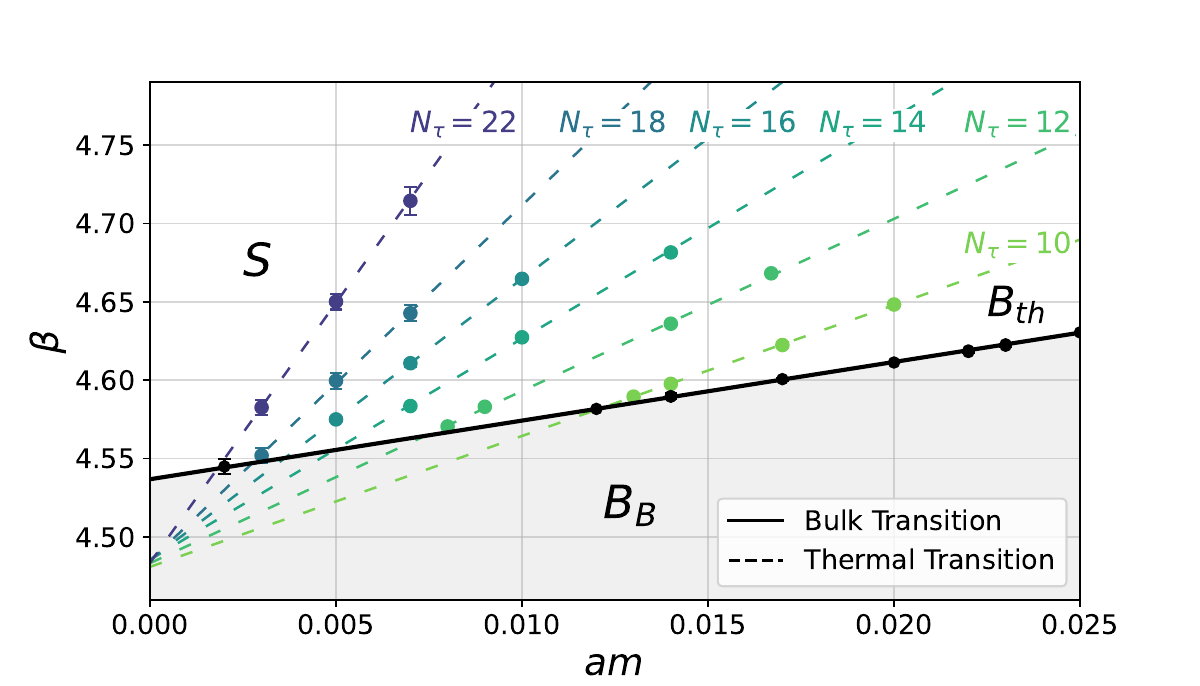}
    }
    \caption{$\Nf=8$}\label{fig:lines-eight}
  \end{subfigure}

  \caption{Chiral phase boundaries for $\Nf=7$ and $8$ in the $(\beta,am)$ plane. Coloured points show the pseudo-critical coupling corresponding to a thermal crossover. The dashed lines are linear fits, extended towards $am=0$. The bulk boundary is $\Nt$-independent and here shown for $\Nt=10$.}
  \label{fig:thermal-lines}
\end{figure}

\section{\texorpdfstring{Comparison of \boldmath$\Nf=7$ and $\Nf=8$}{Comparison of Nf=7 and Nf=8}}
\label{sec:results}

The pseudo-critical hypersurface associated with the chiral transition for QCD with $\Nf=7$ and $\Nf=8$ degenerate flavours is shown in
Figure~\ref{fig:thermal-lines} in the $(\beta,am)$-plane for temporal lattice extents $10\leq \Nt \leq 22$ covering bare masses $0.002\leq am\leq0.025$.
Consistent with previous findings~\cite{Cuteri:2021ikv},
all simulated thermal transitions in this parameter range are compatible with crossovers,
as indicated by kurtosis values $B_4>1.61$
which are growing with increasing volumes\footnote{With increasing spatial volume, the kurtosis approaches $B_4=1.604$ at a $Z(2)$ critical point, $B_4=1$ for a first-order transition, and $B_4=3$ for a crossover.}.
Moreover, all transition lines for fixed $\Nt$ are well described by linear fits
\begin{equation}
\bpc(am,\Nt)=\beta_0(\Nt)+C(\Nt)am.
\label{eq:thermal-lines}
\end{equation}
The bulk phase boundary (solid black line) separates the unphysical bulk phase $B_B$ from the physical regimes $S$ and $B_{\rm th}$. It is identified by remaining nearly identical between
$\Nt=8$ and $\Nt=10$.

\subsection{\texorpdfstring{\boldmath$\Nf=7$}{Nf=7}}

Previous studies~\cite{Cuteri:2021ikv,Klinger:2025xxb,Klinger2026} have shown that the order of the massless chiral transition for $\Nf=7$ depends on $\Nt$: it is first order on coarse lattices and becomes second order only for $\Nt\geq\Nt^\mathrm{tric}$, with current estimates suggesting $\Nt^\mathrm{tric} \approx 23$. A precise determination of $\Nt^\mathrm{tric}$ requires disentangling the thermal and bulk transitions and is still in progress. For $\Nt<\Nt^\mathrm{tric}$, a linear chiral extrapolation is therefore justified at leading order, whereas for $\Nt>\Nt^\mathrm{tric}$ one expects $O(2)$ critical scaling near the chiral limit, reflecting the remnant chiral symmetry of staggered fermions.

The critical couplings $\beta_0(\Nt)$ in the chiral limit are obtained by extrapolating the linear fits of Eq.~\eqref{eq:thermal-lines}, shown in Figure~\ref{fig:lines-seven}, to $am=0$. As shown in Figure~\ref{fig:intercepts-seven}, these couplings increase monotonically with $\Nt$, consistent with a thermal transition connected to the asymptotically free continuum limit, $\beta\to\infty$ as $\Nt\to\infty$.
For $\Nt=18$ and $22$, these intercepts lie above the bulk boundary, so that the thermal transition emerges in the chiral limit on the physical weak-coupling side. Setting a scale, the corresponding intercepts $\beta_0(\Nt)$ represent critical temperatures $T_c(am=0,\Nt)$ for those lattice spacings. This behaviour corresponds to the QCD-like scenario in Figure~\ref{fig:beta-am-qcd}.

\subsection{\texorpdfstring{\boldmath$\Nf=8$}{Nf=8} and conformal hyperscaling}
\label{sec:hyperscaling}

\begin{figure}[tp]
  \centering
  \begin{subfigure}[t]{.5\linewidth}
    \centering
    \makebox[\linewidth][c]{\includegraphics[width=1.10\linewidth]{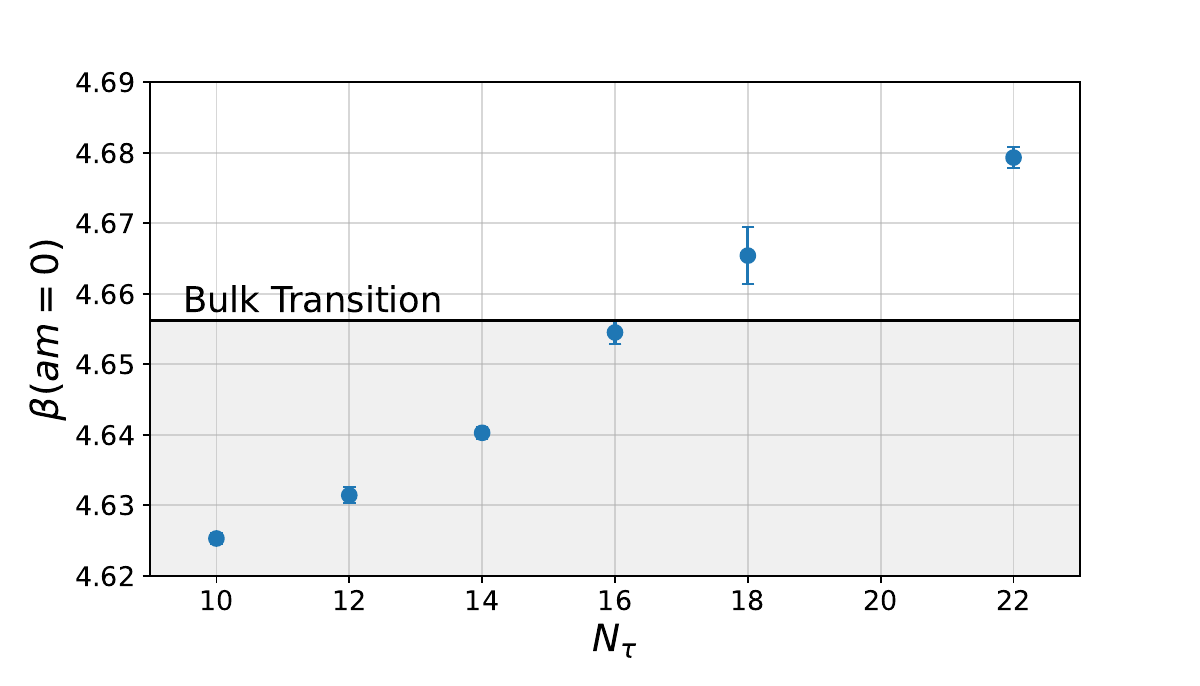}
    }
    \caption{$\Nf=7$}\label{fig:intercepts-seven}
  \end{subfigure}\begin{subfigure}[t]{.5\linewidth}
    \centering
    \makebox[\linewidth][c]{\includegraphics[width=1.10\linewidth]{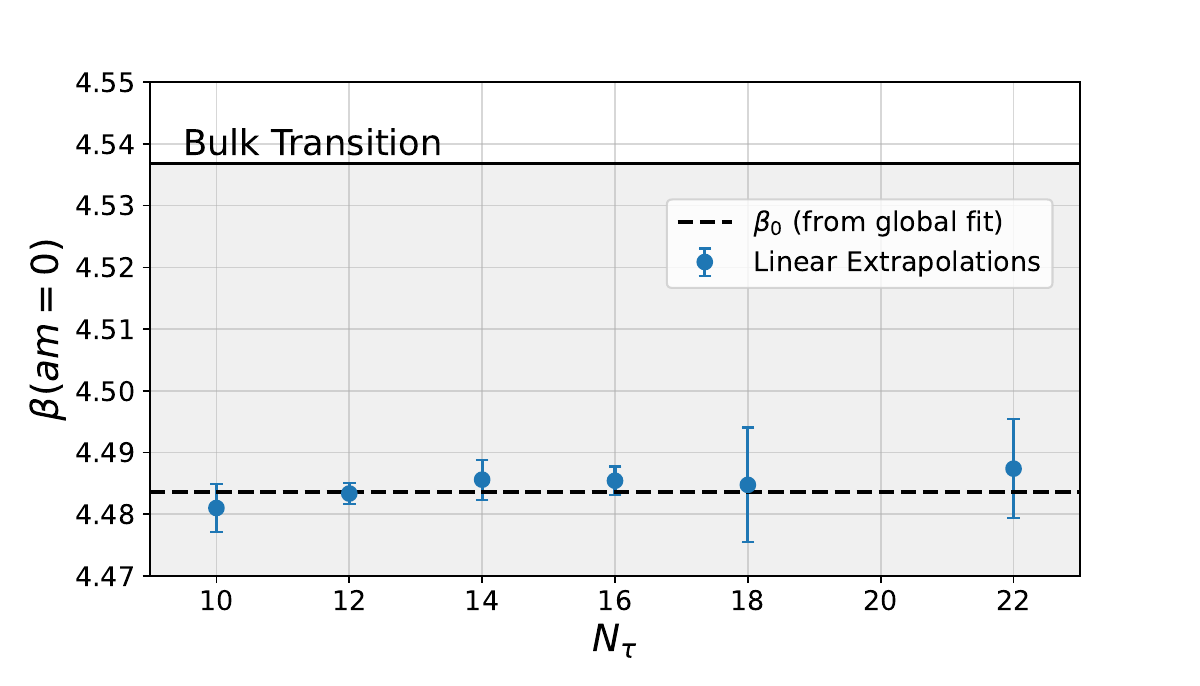}
    }
    \caption{$\Nf=8$}\label{fig:intercepts-eight}
  \end{subfigure}

\caption{Chiral intercepts obtained by extrapolating the thermal lines (Eq.~\eqref{eq:thermal-lines}) to $am=0$, as a function of $\Nt$.
The solid black lines mark the extrapolated bulk boundaries at $am=0$ for each $\Nf$.
For $\Nf=8$, all intercepts are compatible with a common value $\beta_0$ (dashed line), obtained by the global fit of Eq.~\eqref{eq:surface}.}
  \label{fig:intercepts}
\end{figure}

For $\Nf=8$, the chiral extrapolations of the linear fits of Eq.~\eqref{eq:thermal-lines}, shown in Figure~\ref{fig:lines-eight}, reveal a strikingly different pattern: all intercepts lie below the bulk boundary and are compatible with a common, i.e.~$\Nt$-independent constant, see Figure~\ref{fig:intercepts-eight}. Over the range of $\Nt$ studied, we thus find no indication that the thermal boundary reaches $am=0$ on the physical weak-coupling side, consistent with the conformal scenario sketched in Figures~\ref{fig:at-g-conformal} and~\ref{fig:beta-am-conformal}.

Note that a common intercept $\beta_0(\Nt)=\beta_0$ on the vertical axis for $\Nt\rightarrow \infty$
is a necessary condition for infrared conformality. Any moving intercept on the  $\beta$-axis would feature a broken low temperature phase in the chiral limit, and thus be QCD-like. On the other hand, an intercept on the $am$-axis would imply a symmetric phase in the $\beta=0$ strong coupling limit, which is ruled out for $N_f \lesssim 52$~\cite{deForcrand:2012vh}.
However, this condition establishes only the absence of spontaneous chiral symmetry breaking in the continuum. Demonstrating the existence of an infrared fixed point requires additional evidence. For this, the functional form of the full pseudo-critical hypersurface provides a more stringent test. For $\Nf=8$, the thermal phase boundaries are well described by
\begin{equation}
  \boxed{\bpc(am,\Nt)=\beta_0+C\Nt^p\,am}.
  \label{eq:surface}
\end{equation}
The power-law growth of the slopes with $\Nt$ has a natural interpretation in terms of conformal hyperscaling in a mass-deformed conformal theory.
At low energies, the massless theory approaches a scale-invariant regime without a dynamically generated infrared mass scale. A non-zero quark mass is a relevant deformation that explicitly breaks conformal symmetry and, in the absence of other relevant deformations, sets the scale for all infrared observables.
In the infrared scaling regime, all mass-induced scales follow the same leading power law in the renormalised quark mass, known as conformal hyperscaling~\cite{Miransky1999,DelDebbioZwicky}.
For the pseudo-critical temperature, this gives
\begin{equation}
T_{\rm pc}\propto m_R^{1/y_m},
\qquad y_m=1+\gamma_m^*.
\label{eq:continuum-hyperscaling}
\end{equation}
Here $m_R$ is defined at a fixed reference scale, and $y_m$ is the RG scaling exponent of the quark mass, determined by the mass anomalous dimension $\gamma_m^*$ at the infrared fixed point.

Rearranging Eq.~\eqref{eq:surface} at fixed $\beta>\beta_0$ gives the lattice temperature
\begin{equation}
aT_{\rm pc}=\frac{1}{N_\tau}
=\left(\frac{C\,am}{\beta-\beta_0}\right)^{1/p}
=K(\beta)(am)^{1/p},
\qquad
K(\beta)=\left(\frac{C}{\beta-\beta_0}\right)^{1/p}.
\label{eq:thermal-hyperscaling}
\end{equation}
For our staggered fermions, multiplicative mass renormalisation implies $am_R\propto am$ at fixed $\beta$ and reference scale. Comparing Eqs.~\eqref{eq:continuum-hyperscaling} and~\eqref{eq:thermal-hyperscaling} we then identify
\begin{equation}
p=y_m=1+\gamma_m^*
\label{eq:mass-exponent}
\end{equation}
in the infrared hyperscaling regime.
Eq.~\eqref{eq:thermal-hyperscaling} implies that for each fixed $\beta>\beta_0$ the pseudo-critical temperature vanishes in the chiral limit, $T_{\rm pc}\to0$ as $am\to0$, consistent with the absence of a thermal transition in the massless theory.

\begin{figure}[!t]
  \centering
  \begin{subfigure}[b]{.5\linewidth}
    \centering
    \makebox[\linewidth][c]{\includegraphics[height=4.5cm,keepaspectratio]{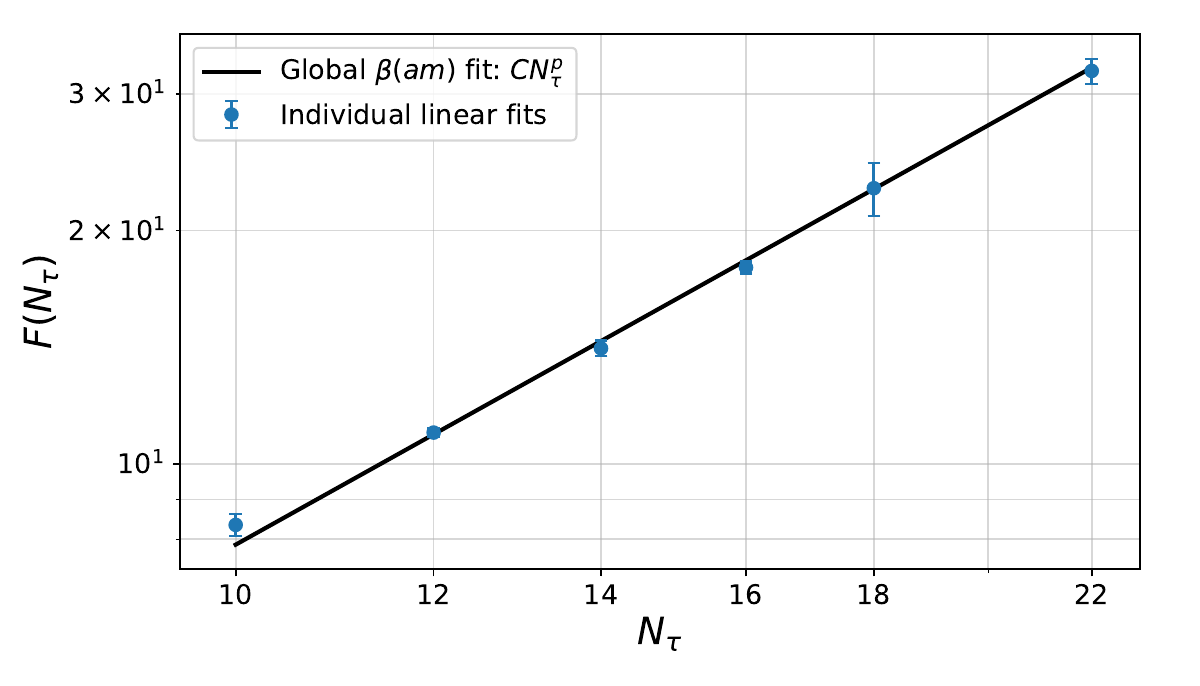}
    }

        \caption{Thermal-line slopes}\label{fig:slopes-power-law}
  \end{subfigure}\begin{subfigure}[b]{.5\linewidth}
    \centering
    \makebox[\linewidth][c]{\raisebox{0.1cm}{\includegraphics[height=4.5cm,keepaspectratio]{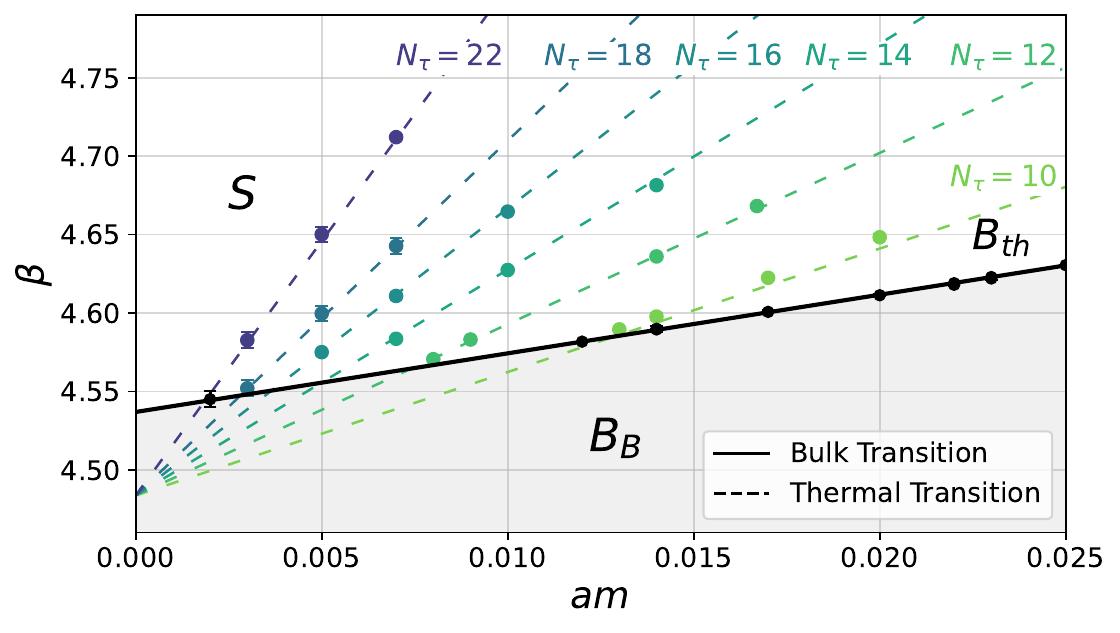}
    } }
    \caption{$\Nf=8$ phase boundaries}\label{fig:surface-boundaries}
  \end{subfigure}
\caption{(\subref{fig:slopes-power-law}) Slopes from the independent linear fits in Figure~\ref{fig:lines-eight}, shown on logarithmic axes.
(\subref{fig:surface-boundaries}) Eight-flavour phase boundaries in the $(\beta,am)$ plane.
The solid line $F(\Nt)=C\Nt^p$ in (\subref{fig:slopes-power-law}) and the thermal lines in (\subref{fig:surface-boundaries}) are obtained from a single global fit of Eq.~\eqref{eq:surface} to the thermal data for $12\leq\Nt\leq22$.}
  \label{fig:slopes}
\end{figure}

Figure~\ref{fig:slopes-power-law} shows that the slopes extracted from the independent linear fits in Figure~\ref{fig:lines-eight} follow a power law in $\Nt$. The solid black line represents $F(\Nt)=C\Nt^p$, obtained from a global fit of Eq.~\eqref{eq:surface} to the thermal data with $12\leq\Nt\leq22$. The corresponding thermal lines are shown in Figure~\ref{fig:surface-boundaries}. With only three parameters, this fit describes the data with $\chi^2/\mathrm{dof}=16.3/13 \simeq1.25$, yielding $\beta_0=4.4836(12)$, $C=0.1260(42)$, and an effective mass anomalous dimension $\gamma_m^\mathrm{eff}=p-1=0.796(11)$.
The fit range was chosen to start at $\Nt=12$, as we found that restricting the fit to progressively larger $N_\tau$ improves the fit quality, while the extracted mass anomalous dimension increases and appears to level off near unity (not shown). We denote it by $\gamma_m^\mathrm{eff}$ to reflect this remaining fit-window dependence. This suggests that corrections to leading-order hyperscaling remain relevant. Such corrections can significantly affect the extracted exponents, as demonstrated in finite-size scaling studies~\cite{Cheng2014}.
A controlled determination of the asymptotic exponent therefore requires an analysis beyond leading order or simulations at larger $\Nt$, while also assessing finite-volume effects.
Our effective values are qualitatively consistent with eight-flavour spectroscopic determinations, which depend appreciably on the observable and fit range~\cite{Appelquist2014,Aoki2017}. Furthermore, our findings agree with simulations using nHYP-smeared staggered fermions at $\Nt=20$ and $24$ which show the same qualitative pattern of the thermal phase boundaries~\cite{Schaich2015,Appelquist2019}.

\section{Conclusions and outlook}
\label{sec:conclusions}

We have investigated the onset of the conformal window by comparing the chiral phase boundaries of seven- and eight-flavour QCD. For $\Nf=7$, the chiral extrapolations indicate a thermal transition consistent with the asymptotically free continuum limit, $\beta\to\infty$ as $\Nt\to\infty$. By contrast, for $\Nf=8$, we find no indication of spontaneous chiral symmetry breaking in the chiral limit on the physical weak-coupling side over the investigated range of $\Nt$.

Our central result is that the eight-flavour thermal hypersurface is well described by
conformal hyperscaling,
$aT_{\rm pc}\propto(am)^{1/p}$, with $p=1+\gamma_m^*$ and $\gamma_m^*$ the mass anomalous dimension at the infrared fixed point. Taken together, the bare parameter phase structure and hyperscaling provide strong evidence that
eight-flavour QCD lies within the conformal window. Whether its onset happens exactly at eight flavours or at a non-integer value between seven and eight remains unresolved at the current precision. Extending the same analysis to non-integer $\Nf$ via a generalised rooted staggered-fermion determinant~\cite{Cuteri:2017gci} offers a way to locate this boundary more precisely.

Our global fit yields an effective mass anomalous dimension $\gamma_m^\mathrm{eff}\simeq0.8$, while fits restricted to larger $\Nt$ yield values closer to unity. This fit-window dependence motivates further tests of corrections to finite-size and
conformal hyperscaling. Finally, a further direction for future research arises from the
observed coupling dependence of the hyperscaling amplitude $K(\beta)\propto(\beta-\beta_0)^{-1/p}$ in Eq.~\eqref{eq:thermal-hyperscaling}.
This behaviour could be explained by a beta-function connecting a non-trivial UV fixed point to the IR fixed point in
the conformal window, which would be particularly intriguing in light of proposals for a distinct strong-coupling phase with symmetric
mass generation at $\Nf=8$~\cite{Hasenfratz2022,HasenfratzWitzel2024}.
Establishing such an interpretation requires independent RG or continuum-scaling tests.

\acknowledgments
\noindent We thank Anna Hasenfratz for helpful discussions. This work is supported by the Deutsche Forschungsgemeinschaft (DFG) through the CRC-TR 211 -- project number 315477589 -- TRR 211. We acknowledge computing resources provided by the HYDRA cluster at GSI Darmstadt and the Goethe-HLR cluster at the Center for Scientific Computing Frankfurt.

\begingroup
\setlength{\bibitemsep}{0.1\baselineskip}
\setlength{\bibparsep}{0pt}
\setlength{\biblabelsep}{0.5em}
\setlength{\bibnamesep}{0pt}
\vspace{-1em}
\defbibheading{bibliography}[\refname]{\section*{\vspace{-1em}#1}}
\printbibliography[title={References}]

@article{BanksZaks,
  author = {Banks, Tom and Zaks, A.},
  title = {{On the phase structure of vector-like gauge theories with massless fermions}},
  journaltitle = {Nucl. Phys. B},
  volume = {196},
  date = {1982},
  pages = {189--204},
  doi = {10.1016/0550-3213(82)90035-9},
}

@article{Hasenfratz2023,
  author = {Hasenfratz, Anna and Neil, Ethan T. and Shamir, Yigal and Svetitsky, Benjamin and Witzel, Oliver},
  title = {{Infrared fixed point of the SU(3) gauge theory with $N_f=10$ flavors}},
  journaltitle = {Phys. Rev. D},
  volume = {108},
  date = {2023},
  eid = {L071503},
  doi = {10.1103/PhysRevD.108.L071503},
  eprint = {2306.07236},
  eprinttype = {arxiv},
  eprintclass = {hep-lat},
}

@article{PetersonHasenfratz2024,
  author = {Hasenfratz, Anna and Peterson, Curtis T.},
  title = {{Infrared fixed point in the massless twelve-flavor SU(3) gauge-fermion system}},
  journaltitle = {Phys. Rev. D},
  volume = {109},
  date = {2024},
  eid = {114507},
  doi = {10.1103/PhysRevD.109.114507},
  eprint = {2402.18038},
  eprinttype = {arxiv},
  eprintclass = {hep-lat},
}

@article{KotovLombardoTrunin2021,
  author = {Kotov, A. {Yu.} and Lombardo, Maria Paola and Trunin, Anton},
  title = {{Gliding down the QCD transition line, from $N_f=2$ till the onset of conformality}},
  journaltitle = {Symmetry},
  volume = {13},
  date = {2021},
  eid = {1833},
  doi = {10.3390/sym13101833},
  eprint = {2111.00569},
  eprinttype = {arxiv},
  eprintclass = {hep-lat},
}

@misc{PastorGutierrez2026,
  author = {Pastor-Gutiérrez, Álvaro},
  title = {{Non-perturbative news from the conformal window}},
  date = {2026},
  eprint = {2608.23674},
  eprinttype = {arxiv},
  eprintclass = {hep-th},
}

@article{Ciambriello2024,
  author = {Ciambriello, Luca and Contino, Roberto and Luzio, Andrea and Romano, Marcello and Xu, Ling-Xiao},
  title = {{A novel strategy to prove chiral symmetry breaking in QCD-like theories}},
  journaltitle = {Phys. Lett. B},
  volume = {862},
  date = {2025},
  eid = {139367},
  doi = {10.1016/j.physletb.2025.139367},
  eprint = {2404.02967},
  eprinttype = {arxiv},
  eprintclass = {hep-th},
}

@article{Klinger2026,
  author = {Klinger, Jan Philipp and Kaiser, Reinhold and Philipsen, Owe and Schaible, Jonas},
  title = {{On the phase structure of massless many-flavour QCD with staggered fermions}},
  journaltitle = {PoS},
  volume = {518},
  issuetitle = {LATTICE2025},
  date = {2026},
  eid = {119},
  doi = {10.22323/1.518.0119},
  eprint = {2603.20099},
  eprinttype = {arxiv},
  eprintclass = {hep-lat},
}

@article{Brown1992,
  author = {Brown, Frank R. and Chen, Hong and Christ, Norman H. and Dong, Zhihua and Mawhinney, Robert D. and Schaffer, Wendy and Vaccarino, Alessandro},
  title = {{Lattice QCD with eight light quark flavors}},
  journaltitle = {Phys. Rev. D},
  volume = {46},
  date = {1992},
  pages = {5655--5670},
  doi = {10.1103/PhysRevD.46.5655},
  eprint = {hep-lat/9206001},
  eprinttype = {arxiv},
  eprintclass = {hep-lat},
}

@article{Iwasaki2004,
  author = {Iwasaki, Y. and Kanaya, K. and Kaya, S. and Sakai, S. and Yoshie, T.},
  title = {{Phase structure of lattice QCD for general number of flavors}},
  journaltitle = {Phys. Rev. D},
  volume = {69},
  date = {2004},
  eid = {014507},
  doi = {10.1103/PhysRevD.69.014507},
  eprint = {hep-lat/0309159},
  eprinttype = {arxiv},
  eprintclass = {hep-lat},
}

@article{CL2QCD,
  author = {Philipsen, Owe and Pinke, Christopher and Sciarra, Alessandro and Bach, Matthias},
  title = {{CL2QCD -- Lattice QCD based on OpenCL}},
  journaltitle = {PoS},
  volume = {214},
  issuetitle = {LATTICE2014},
  date = {2014},
  eid = {038},
  doi = {10.22323/1.214.0038},
  eprint = {1411.5219},
  eprinttype = {arxiv},
  eprintclass = {hep-lat},
}

@article{BaHaMAS,
  author = {Sciarra, Alessandro},
  title = {{BaHaMAS: A Bash Handler to Monitor and Administrate Simulations}},
  journaltitle = {EPJ Web Conf.},
  volume = {175},
  date = {2018},
  eid = {09003},
  doi = {10.1051/epjconf/201817509003},
  eprint = {1710.08831},
  eprinttype = {arxiv},
  eprintclass = {hep-lat},
}

@article{FerrenbergSwendsen1989,
  author = {Ferrenberg, Alan M. and Swendsen, Robert H.},
  title = {{Optimized Monte Carlo data analysis}},
  journaltitle = {Phys. Rev. Lett.},
  volume = {63},
  date = {1989},
  pages = {1195--1198},
  doi = {10.1103/PhysRevLett.63.1195},
}

@article{Miransky1999,
  author = {Miransky, V. A.},
  title = {{Dynamics in the conformal window in QCD-like theories}},
  journaltitle = {Phys. Rev. D},
  volume = {59},
  date = {1999},
  eid = {105003},
  doi = {10.1103/PhysRevD.59.105003},
  eprint = {hep-ph/9812350},
  eprinttype = {arxiv},
  eprintclass = {hep-ph},
}

@article{DelDebbioZwicky,
  author = {Del Debbio, Luigi and Zwicky, Roman},
  title = {{Hyperscaling relations in mass-deformed conformal gauge theories}},
  journaltitle = {Phys. Rev. D},
  volume = {82},
  date = {2010},
  eid = {014502},
  doi = {10.1103/PhysRevD.82.014502},
  eprint = {1005.2371},
  eprinttype = {arxiv},
  eprintclass = {hep-ph},
}

@article{Appelquist2014,
  author = {Appelquist, T. and Brower, R. C. and Fleming, G. T. and Kiskis, J. and Lin, M. F. and Neil, E. T. and Osborn, J. C. and Rebbi, C. and Rinaldi, E. and Schaich, D. and Schroeder, C. and Syritsyn, S. and Voronov, G. and Vranas, P. and Weinberg, Evan and Witzel, O.},
  title = {{Lattice simulations with eight flavors of domain wall fermions in SU(3) gauge theory}},
  journaltitle = {Phys. Rev. D},
  volume = {90},
  date = {2014},
  eid = {114502},
  usera = {LSD Collaboration},
  doi = {10.1103/PhysRevD.90.114502},
  eprint = {1405.4752},
  eprinttype = {arxiv},
  eprintclass = {hep-lat},
}

@article{Aoki2017,
  author = {Aoki, Yasumichi and Aoyama, Tatsumi and Bennett, E. and Kurachi, Masafumi and Maskawa, Toshihide and Miura, Kohtaroh and Nagai, Kei-ichi and Ohki, Hiroshi and Rinaldi, Enrico and Shibata, Akihiro and Yamawaki, Koichi and Yamazaki, Takeshi},
  title = {{Light flavor-singlet scalars and walking signals in $N_f=8$ QCD on the lattice}},
  journaltitle = {Phys. Rev. D},
  volume = {96},
  date = {2017},
  eid = {014508},
  usera = {LatKMI Collaboration},
  doi = {10.1103/PhysRevD.96.014508},
  eprint = {1610.07011},
  eprinttype = {arxiv},
  eprintclass = {hep-lat},
}

@article{Hasenfratz2022,
  author = {Hasenfratz, Anna},
  title = {{Emergent strongly coupled ultraviolet fixed point in four dimensions with 8 K\"ahler-Dirac fermions}},
  journaltitle = {Phys. Rev. D},
  volume = {106},
  date = {2022},
  eid = {014513},
  doi = {10.1103/PhysRevD.106.014513},
  eprint = {2204.04801},
  eprinttype = {arxiv},
  eprintclass = {hep-lat},
}

@article{HasenfratzWitzel2024,
  author = {Witzel, Oliver and Hasenfratz, Anna},
  title = {{Investigating SU(3) with $N_f=8$ fundamental fermions at strong renormalized coupling}},
  journaltitle = {PoS},
  volume = {466},
  issuetitle = {LATTICE2024},
  date = {2025},
  eid = {146},
  usera = {Lattice Strong Dynamics Collaboration},
  doi = {10.22323/1.466.0146},
  eprint = {2412.10322},
  eprinttype = {arxiv},
  eprintclass = {hep-lat},
}

@article{Braun:2006jd,
    author = "Braun, Jens and Gies, Holger",
    title = "{Chiral phase boundary of QCD at finite temperature}",
    eprint = "hep-ph/0602226",
    archivePrefix = "arXiv",
    doi = "10.1088/1126-6708/2006/06/024",
    journal = "JHEP",
    volume = "06",
    pages = "024",
    year = "2006"
}

@article{Braun:2009ns,
    author = "Braun, Jens and Gies, Holger",
    title = "{Scaling laws near the conformal window of many-flavor QCD}",
    eprint = "0912.4168",
    archivePrefix = "arXiv",
    primaryClass = "hep-ph",
    doi = "10.1007/JHEP05(2010)060",
    journal = "JHEP",
    volume = "05",
    pages = "060",
    year = "2010"
}

@article{Cuteri:2021ikv,
    author = "Cuteri, Francesca and Philipsen, Owe and Sciarra, Alessandro",
    title = "{On the order of the QCD chiral phase transition for different numbers of quark flavours}",
    eprint = "2107.12739",
    archivePrefix = "arXiv",
    primaryClass = "hep-lat",
    doi = "10.1007/JHEP11(2021)141",
    journal = "JHEP",
    volume = "11",
    pages = "141",
    year = "2021"
}

@article{Klinger:2025xxb,
    author = "Klinger, Jan Philipp and Kaiser, Reinhold and Philipsen, Owe",
    title = "{The order of the chiral phase transition in massless many-flavour lattice QCD}",
    eprint = "2501.19251",
    archivePrefix = "arXiv",
    primaryClass = "hep-lat",
    doi = "10.22323/1.466.0172",
    journal = "PoS",
    volume = "LATTICE2024",
    pages = "172",
    year = "2025"
}

@article{Cuteri:2017gci,
    author = "Cuteri, Francesca and Philipsen, Owe and Sciarra, Alessandro",
    title = "{QCD chiral phase transition from noninteger numbers of flavors}",
    eprint = "1711.05658",
    archivePrefix = "arXiv",
    primaryClass = "hep-lat",
    doi = "10.1103/PhysRevD.97.114511",
    journal = "Phys. Rev. D",
    volume = "97",
    number = "11",
    pages = "114511",
    year = "2018"
}

@article{deForcrand:2012vh,
    author = "de Forcrand, Philippe and Kim, Seyong and Unger, Wolfgang",
    title = "{Conformality in many-flavour lattice QCD at strong coupling}",
    eprint = "1208.2148",
    archivePrefix = "arXiv",
    primaryClass = "hep-lat",
    reportNumber = "CERN-PH-TH-2012-204",
    doi = "10.1007/JHEP02(2013)051",
    journal = "JHEP",
    volume = "02",
    pages = "051",
    year = "2013"
}

@article{Cheng2014,
  author        = {Cheng, Anqi and Hasenfratz, Anna and Liu, Yuzhi
                   and Petropoulos, Gregory and Schaich, David},
  title         = {Finite size scaling of conformal theories in the presence of a near-marginal operator},
  journal       = {Phys. Rev. D},
  volume        = {90},
  pages         = {014509},
  year          = {2014},
  doi           = {10.1103/PhysRevD.90.014509},
  eprint        = {1401.0195},
  archivePrefix = {arXiv},
  primaryClass  = {hep-lat}
}

@inproceedings{Schaich2015,
  author        = {Schaich, David and Hasenfratz, Anna and Rinaldi, Enrico},
  collaboration = {Lattice Strong Dynamics},
  title         = {Finite-temperature study of eight-flavor {SU(3)} gauge theory},
  booktitle     = {Origin of Mass and Strong Coupling Gauge Theories ({SCGT15})},
  publisher     = {World Scientific},
  pages         = {351--354},
  year          = {2018},
  doi           = {10.1142/9789813231467_0051},
  eprint        = {1506.08791},
  archivePrefix = {arXiv},
  primaryClass  = {hep-lat}
}

@article{Appelquist2019,
  author        = {Appelquist, Thomas and Brower, Richard C. and Fleming, George T.
                   and Gasbarro, Andrew and Hasenfratz, Anna and Jin, Xiao-Yong
                   and Neil, Ethan T. and Osborn, James C. and Rebbi, Claudio
                   and Rinaldi, Enrico and Schaich, David and Vranas, Pavlos
                   and Weinberg, Evan and Witzel, Oliver},
  collaboration = {Lattice Strong Dynamics},
  title         = {Nonperturbative investigations of {SU(3)} gauge theory with eight dynamical flavors},
  journal       = {Phys. Rev. D},
  volume        = {99},
  pages         = {014509},
  year          = {2019},
  doi           = {10.1103/PhysRevD.99.014509},
  eprint        = {1807.08411},
  archivePrefix = {arXiv},
  primaryClass  = {hep-lat}
}

@article{MiuraLombardo2013,
  author       = {Miura, Kohtaroh and Lombardo, Maria Paola},
  title        = {Lattice Monte-Carlo study of pre-conformal dynamics in strongly flavoured {QCD} in the light of the chiral phase transition at finite temperature},
  journaltitle = {Nucl. Phys. B},
  volume       = {871},
  date         = {2013},
  pages        = {52--81},
  doi          = {10.1016/j.nuclphysb.2013.02.008},
  eprint       = {1212.0955},
  eprinttype   = {arxiv},
  eprintclass  = {hep-lat},
}

@article{DElia2026,
  author       = {D'Elia, Massimo and Nacci, Marco and Zambello, Kevin},
  title        = {{Roberge--Weiss} transition as a probe for conformality in many-flavor {QCD}},
  journaltitle = {Phys. Rev. D},
  volume       = {114},
  date         = {2026},
  eid          = {034502},
  doi          = {10.1103/d8jw-ydq7},
  eprint       = {2604.07284},
  eprinttype   = {arxiv},
  eprintclass  = {hep-lat},
}

@article{Lombardo_12,
  title = {Evidence for a conformal phase in {$\mathrm{SU}(N)$} gauge theories},
  author = {Deuzeman, A. and Lombardo, M. P. and Pallante, E.},
  journal = {Phys. Rev. D},
    eprint = "0904.4662",
    archivePrefix = "arXiv",
  volume = {82},
  issue = {7},
  pages = {074503},
  numpages = {11},
  year = {2010},
  month = {Oct},
  publisher = {American Physical Society},
  doi = {10.1103/PhysRevD.82.074503},
  url = {https://link.aps.org/doi/10.1103/PhysRevD.82.074503}
}
\endgroup

\end{document}